\documentclass{epl2}

\newcommand{\wo}{\mbox{$\omega_0$}}
\newcommand{\wk}{\mbox{$\omega_k$}}
\newcommand{\wkp}{\mbox{$\omega_{k'}$}}
\newcommand{\akj}{\mbox{$a_{{\bf k}j}$}}
\newcommand{\adkj}{\mbox{$a_{{\bf k}j}^\dagger$}}
\newcommand{\gkji}{\mbox{$\epsilon^{(i)}_{{\bf k}j}$}}

\newcommand{\gkja}{\mbox{$\epsilon^{(A)}_{{\bf k}j}$}}
\newcommand{\gkjb}{\mbox{$\epsilon^{(B)}_{{\bf k}j}$}}
\newcommand{\gkpja}{\mbox{$\epsilon^{(A)}_{{\bf k}'j'}$}}
\newcommand{\gkpjb}{\mbox{$\epsilon^{(B)}_{{\bf k}'j'}$}}
\newcommand{\epki}{\mbox{$e^{i{\bf k}\cdot {\bf R}_i}$}}
\newcommand{\emki}{\mbox{$e^{-i{\bf k}\cdot {\bf R}_i}$}}
\newcommand{\ekj}{\mbox{$\hat{\bf e}_{{\bf k}j}$}}
\newcommand{\bk}{\mbox{${\bf k}$}}
\newcommand{\bR}{\mbox{${\bf R}$}}
\newcommand{\ketggv}{\mbox{$\mid gg \{ 0_{{\bf k} j} \} \rangle$}}
\newcommand{\keteev}{\mbox{$\mid ee \{0_{{\bf k} j}\} \rangle$}}
\newcommand{\ketegk}{\mbox{$\mid eg 1_{{\bf k} j} \rangle$}}
\newcommand{\ketgek}{\mbox{$\mid ge 1_{{\bf k} j} \rangle$}}

\newcommand{\ketggkk}{\mbox{$\mid gg 1_{{\bf k} j}1_{{\bf k}'j'} \rangle$}}
\newcommand{\keteekk}{\mbox{$\mid ee 1_{{\bf k} j}1_{{\bf k}'j'} \rangle$}}

\def\bk{{\bf k}}

\def\bR{{\bf R}}

\def\w{\omega}
\def\wk{\omega_k}
\def\w0{\omega_0}
\def\ekj{\hat{\bf e}_{\bk j}}

\def\ekj{\hat{e}_{\bk j}}

\newcommand{\ketbra}[2]{|{#1}\rangle\langle{#2}|}       
\newcommand{\ave}[1]{\langle{#1}\rangle}            
\newcommand{\ket}[1]{|{#1}\rangle}
\newcommand{\bra}[1]{\langle{#1}|}

\title{Casimir - Polder potentials as entanglement probe.}
\author{M.A.Cirone \inst{1}, G. Compagno \inst{1}, G. M. Palma \inst{2}, R. Passante \inst{1}, \and F. Persico \inst{1} }
\institute{
  \inst{1}CNISM \& Dipartimento di Scienze Fisiche ed Astronomiche, Universit\`a degli Studi di Palermo,  via Archirafi 36, I-90123 Palermo (Italy)\\
  \inst{2} NEST - CNR (INFM) \& Dipartimento di Scienze Fisiche ed Astronomiche, Universit\`a degli Studi di Palermo,  via Archirafi 36, I-90123 Palermo
  (Italy)}
  \pacs{03.65.Ud}{ Entanglement and quantum nonlocality }
  \pacs{03.67.Mn}{ Entanglement production,
characterization, and manipulation} \pacs{42.50.Dv}{ Nonclassical
states of the electromagnetic field, including entangled photon
states; quantum state engineering and measurements}
 \abstract{We
have considered the interaction of a pair of spatially separated
two-level atoms with the electromagnetic field in its vacuum state
and we have analyzed the amount of entanglement induced between
the two atoms by the non local field fluctuations. This has
allowed us to characterize the quantum nature of the non local
correlations of the electromagnetic field vacuum state as well as
to link the induced quantum entanglement with Casimir - Polder
potentials.}

\begin{document}

\maketitle

The zero - point fluctuations of the vacuum state of the electromagnetic field are characterized by strong non local correlations
\cite{Milonni,Fearn,book} which are at the origin of phenomena like Casimir - Polder forces \cite{Thiru,Casimir}. An interesting
open problem is the possibility to characterize the quantum nature of such non local correlations. So far most of the efforts done
towards this direction have focused on violations of suitable Bell's inequalities by   the vacuum state fluctuations
\cite{Werner}. However a direct experimental detection of such inequalities violation for the vacuum state is awkward. In this
letter we shall take a somewhat different approach. It is a well known fact that when two quantum subsystems, e.g. two atoms,
interact with a common bath, they become entangled (see for instance \cite{Braun}). Such a pair of subsystems can therefore be
used as a probe of the non local vacuum field fluctuations. In other words the quantum nature of such fluctuations can be
characterized by the amount of entanglement induced between the two spatially separated probe atoms. Some interesting
contributions in this direction have already appeared in literature \cite{Reznik}. In the following we shall quantify the
entanglement induced between the two probe atoms by means of the concurrence \cite{Wootters} as this is amenable to a
straightforward physical interpretation \cite{Cirone}. Indeed we will show that the concurrence turns out to be linked to the
Casimir - Polder potentials. Casimir-Polder forces are long-range interactions between neutral atoms or molecules arising from
their interaction with the common electromagnetic radiation field in its vacuum state. For atoms in the ground state the
Casimir-Polder potential behaves as $R^{-6}$ for interatomic distances smaller than a characteristic distance of the order of an
appropriate average of the atomic transition wavelengths  -  but large enough to neglect any overlap between the electron
wavefunctions - and as $R^{-7}$ for larger distances.

Casimir-Polder potentials have been experimentally detected in several physical systems.  Here we mention two experiments, which are closer to our analysis \cite{hinds1,hinds2}, in which the deflection of neutral atoms due to the interaction  with conducting plates, has been measured.
It is possible to show that the Casimir-Polder potential is also strictly
related to the spatial correlations of the vacuum fluctuations of the electromagnetic field.
In fact, it can be obtained from the
classical interaction of the instantaneous dipole moments of the two atoms, which are induced by the spatially correlated
zero-point quantum field fluctuations \cite{Lucia}. Furthermore the use of (three-body) dynamical Casimir-Polder forces for
investigating the nonlocality of field correlations has been also recently suggested \cite{Lucia2}

To introduce all the mathematical tools used in our analysis let us briefly recall the definition and basic properties of the
concurrence. Quantum Information Theory has provided new powerful mathematical tools to quantify the amount of entanglement
between quantum subsystems in some specific situations. In particular for the case of a mixed state of two two-level systems the
entanglement of formation ${\cal E}_{F}$ is a suitable entanglement quantifier \cite{Wootters}, as it quantifies the amount of non
local resources needed to create a given state. From a mathematical viewpoint the main advantage of considering the entanglement
of formation is the fact that it is a monotone function of the so - called concurrence, a quantity which is relatively easy to
calculate. For an arbitrary bipartite system described by the density operator $\rho$, the entanglement of formation ${\cal
E}_{F}$ turns out to be equal to ~\cite{Wootters}

\begin{equation}
{\cal E}_{F}(\rho)=-x\log_{2}x-(1-x)\log_{2}(1-x)
\end{equation}
where $x=(1+\sqrt{1-C^2(\rho)})/2$ and the concurrence $C(\rho)$ is defined as

\begin{equation}
C(\rho)=\max\left\{0,\alpha_{1}-\alpha_{2}-{\alpha_{3}}-{\alpha_{4}}\right\}.
\end{equation}
where $\left\{\alpha_{i}\right\}$ ($i=1,..,4$) are the square
roots of the eigenvalues (in non-increasing order) of the
non-Hermitian operator
$\bar{\rho}=\rho(\sigma_{y}\otimes\sigma_{y})\rho^{*}(\sigma_{y}\otimes\sigma_{y})$,
$\sigma_{y}$ is the y-Pauli operator and $\rho^{*}$ is the complex
conjugate of $\rho$, in the basis of $\sigma_z$ operator. Since
the entanglement of formation is a monotonic function of the
concurrence we will use the latter as entanglement quantifier.
Furthermore, as anticipated, the concurrence turns out to be
directly linked to the Casimir - Polder potential, and therefore
amenable of an experimental detection.

The system we have considered consists of a pair of spatially
separated two-level atoms, A and B, placed at ${\bf R}_A$ and
${\bf R}_B$ respectively, at a distance ${\bf R} \equiv {\bf R}_A
- {\bf R}_B$ large enough to neglect any overlap between the
electron wavefunc{\bf t}ions, interacting with the the
electromagnetic field in its vacuum state. The atom--radiation
system is described by the multipolar Hamiltonian in the dipole
approximation

\begin{equation}
H=H_{AB}+H_F+H_{AB,F}
\end{equation}

with

\begin{eqnarray}
H_{AB} & = & \hbar \wo \sum_{i=A,B} S_z^{(i)} \\
H_F & = & \sum_{{\bf k}j} \hbar \wk \adkj \akj \\
H_{AB, F} & = & \sum_{i=A,B}\sum_{{\bf k}j}
\left[ \left(\gkji \akj \epki S_+^{(i)} +\gkji^*\adkj \emki S_-^{(i)}
 \right) \right. \nonumber \\
& & \left. - \left( \gkji \adkj \emki S_+^{(i)} +\gkji^*\akj \epki
S_-^{(i)} \right)  \right]
\end{eqnarray}
where $\wo=c\,k_0$ represents the separation in angular frequency
of the two atomic levels, $S_{z}$, $S_{+}$ and $S_{-}$ represent
the atomic pseudo--spin operators, $\akj$ and $\adkj$ denote the
annihilation and creation operators, respectively, of photons with
wavevector $\bf{k}$ and polarization $j$, and the coupling
constant, in the multipolar representation,

\begin{equation}
\gkji\equiv i\sqrt{\frac{2\pi \hbar \wk }{V}} \ekj \cdot {\bf
d}^{(i)}
\end{equation}
is purely imaginary for real atomic dipoles ${\bf d}^{(i)}$ and linear polarization unit vectors $\ekj$ (note that no rotating
wave approximation has been made). The two atomic states considered might be, for example, the 1s ground state and one of the 2p
excited states of a hydrogen atom, which satisfy the electric dipole selection rules for the emission of a photon. In this case,
all the three degenerate p excited states equally participate to the interaction with the field, of course. However, in the
following we shall consider the interaction of the ground state with only one of these states, because this makes easier the
connection of our results for the concurrence with the Casimir-Polder forces. On the other hand, a sum over the three degenerate
excited states can be done at the end of the calculation, restoring the spherical symmetry of the atoms; in fact, it is known from
the theory of Casimir-Polder forces that the relevant virtual transitions to the excited states occur independently, up to the
required order in perturbation theory. This makes the two-level system considered here a convenient system to obtain and analyze
the physically relevant quantities.


The normalized, dressed ground state of the two atoms can be
written at second order of approximation in the electric dipole
$O(d^2)$ in the compact form

\begin{eqnarray}
| \psi \rangle & = & c_{gg} \ketggv +\sum_{{\bf k} j}c_{eg,{\bf k}
j} \ketegk
+\sum_{{\bf k} j}c_{ge,{\bf k} j} \ketgek+c_{ee}\keteev \nonumber \\
& & + \frac{1}{2}\sum_{{\bf k} j}\sum_{{\bf k}'j'}c_{gg,{\bf k} j{\bf k}'j'}\ketggkk
+ \frac{1}{2}\sum_{{\bf k} j}\sum_{{\bf k}'j'}c_{ee,{\bf k} j{\bf k}'j'}\keteekk
\label{dress}
\end{eqnarray}
where $\ket{g}$ and $\ket{e}$ denote, respectively, the ground and
excited state of each atom. The explicit form of the various
probability amplitudes will be given later on. We remark the
different properties of the first- and second-order corrections
(i.e., the terms with odd and even number of photons,
respectively): whereas the former give local effects, such as the
Lamb shift of the energy levels of each atom, the latter are
responsible for the Casimir-Polder interaction and are thus
essentially of non local character.

The reduced density operator of the two atoms $\varrho_{AB} = Tr_{field} \ket{\psi}\bra{\psi}$, obtained by tracing over the field
variables, has nonvanishing entries only in the two diagonals of its matrix representation. For this class of density operators
the concurrence $C$ can be expressed in terms of physical quantities, such as correlation functions and average values
\cite{Palma}:
\begin{eqnarray}
C & = & 2{\rm max} \{ 0, C_1, C_2\}, \\
C_1 & \equiv &
\sqrt{(g_{xx}-g_{yy})^2+(g_{xy}+g_{yx})^2}-\sqrt{\left(\frac{1}{4}-g_{zz}
\right)^2-\delta S_z^2}, \\
C_2 & \equiv &
\sqrt{(g_{xx}+g_{yy})^2+(g_{xy}-g_{yx})^2}-\sqrt{\left(\frac{1}{4}+g_{zz}
\right)^2-M_z^2}
\end{eqnarray}
where $g_{ij} $ is defined in terms of expectation value of spin operators as
\begin{eqnarray}
g_{ij} \equiv \langle S_{i}^A S_{j}^B \rangle, & M_i \equiv
\langle S_i^A+S_i^B \rangle/2, & \delta S_i \equiv \langle
S_i^A-S_i^B \rangle
\end{eqnarray}
Such quantities can be evaluated with the help of the wave function eq.(\ref{dress}) therefore dispensing us from an explicit
calculation of the density operator $\varrho_{AB}$. Straightforward algebraic manipulations give a simpler form for $C_1$ and
$C_2$ for our two-atom system:
\begin{eqnarray}
C_1 & \equiv & \sqrt{\ave{\ketbra{ee}{gg}}\ave{\ketbra{gg}{ee}}}
-\sqrt{\ave{\ketbra{eg}{eg}}\ave{\ketbra{ge}{ge}}} \label{c1}\\
C_2 & \equiv & \sqrt{\ave{\ketbra{eg}{ge}}\ave{\ketbra{ge}{eg}}}
-\sqrt{\ave{\ketbra{ee}{ee}}\ave{\ketbra{gg}{gg}}} \label{c2}
\end{eqnarray}

Our goal is the evaluation of the concurrence at second order of
approximation. Since $\ave{\ketbra{gg}{gg}}=1+O(d^2)$, the average
value $\ave{\ketbra{ee}{ee}}$ must be evaluated at fourth order of
approximation. From eq.(\ref{dress}) we obtain
\begin{eqnarray}
\ave{\ketbra{ee}{gg}} & = & \ave{\ketbra{gg}{ee}}^* = \; c_{gg}c^*_{ee}, \\
\ave{\ketbra{eg}{ge}} & = & \ave{\ketbra{ge}{eg}}^* = \;
\sum_{\bk j}c_{eg,\bk j}\; c^*_{ge,\bk j}, \\
\ave{\ketbra{eg}{eg}} & = & \sum_{\bk j} |c_{eg,\bk j}|^2,
\;\;\;\;\; \ave{\ketbra{ge}{ge}} \;\; = \;\; \sum_{\bk j}
|c_{ge,\bk j}|^2, \\
\ave{\ketbra{ee}{ee}} & = & |c_{ee}|^2+\frac{1}{2} \sum_{\bk,j}
\sum_{\bk',j'}|c_{ee,\bk j \bk' j'}|^2
\end{eqnarray}
The probability amplitudes required for the evaluation of the
concurrence are
\begin{eqnarray}
c_{eg,\bk j} & = & -\frac{\gkja e^{-i\bk \cdot \bR_A
}}{\hbar(\wo+\wk)}, \\
c_{ge,\bk j} & = & -\frac{\gkjb e^{-i\bk \cdot \bR_B
}}{\hbar(\wo+\wk)}, \\
 c_{ee} & = & -\frac{1}{2\hbar \wo}\sum_{\bk
j}\frac{\gkja \gkjb (e^{i\bk \cdot \bR }+e^{-i\bk \cdot \bR
})}{\hbar(\wo+\wk)} \\
 c_{ee, \bk j \bk ' j'} & = & \frac{\gkja \gkpjb e^{-i\bk \cdot \bR_A }e^{-i\bk ' \cdot
 \bR_B
}}{\hbar^2(\wo+\wk)(\wo+\wkp)} + (A \longleftrightarrow B)
\end{eqnarray}
With the help of these expressions eq.(\ref{c2}) becomes

\begin{eqnarray}
C_2 & = & \left[ \sum_{\bk j} \frac{\gkja \gkjb e^{-i\bk \cdot \bR
}}{\hbar^2(\wo+\wk)^2} \sum_{\bk ' j'} \frac{\gkpja \gkpjb e^{i\bk
'
\cdot \bR }}{\hbar^2(\wo+\wkp)^2}\right]^{1/2} \nonumber \\
& & -\left[ \sum_{\bk j} \frac{\gkja \gkjb e^{-i\bk \cdot \bR
}}{\hbar^2(\wo+\wk)^2} \sum_{\bk ' j'} \frac{\gkpja \gkpjb e^{i\bk
'
\cdot \bR }}{\hbar^2(\wo+\wkp)^2} \right. \nonumber \\
& & \left. + \sum_{\bk j}\frac{|\gkja|^2 }{\hbar^2(\wo+\wk)^2}
\sum_{\bk ' j'} \frac{|\gkpjb|^2 }{\hbar^2(\wo+\wkp)^2} +
|c_{ee}|^2 \right]^{1/2}
\end{eqnarray}
Thus $C_2$ is negative and the concurrence is $C={\rm max}(0,2 \,
C_1)$, with

\begin{equation}
C_1 = |c_{ee}|-\left[\sum_{\bk j}c_{eg,\bk j}\sum_{\bk '
j'}c_{ge,\bk ' j'}\right]^{1/2}
\end{equation}
The sums in square brackets formally diverge when the sum runs from $k=0$ to $k\rightarrow \infty$; however, as noted above, the
probability amplitudes $c_{eg,\bk j}$ and $c_{ge,\bk j}$ describe local interactions of each atom with the vacuum field and play
no role in the interaction between the two atoms. Such terms are responsible only  for the appearance of local energy shifts and
do not contribute to interatomic forces nor to non local correlation of the fluctuations of atomic dipoles. Indeed, in the
research literature on the interaction of pairs of atoms with vacuum fluctuations, all such terms which do not depend on the
atomic separation (and therefore diverge) are neglected - see e.g.\cite{Milonni, Berman, Power} . Following this approach the
concurrence is

\begin{equation}
C= 2 \, C_1 = 2|c_{ee}|
\end{equation}
Therefore for any separation $R$ the concurrence, and thus the
entanglement, is determined by $c_{ee}$ only. This fact is
consistent with the observation that the term $c_{ee}\keteev$ in
the dressed ground state Eq.(\ref{dress}) is entirely due to the
interaction between the two atoms via the common electromagnetic
field.

When we transform the sums over the wave vector ${\bf k}$ into
integrals following the usual prescription $V^{-1}\sum_{\bk}
\rightarrow (2\pi)^{-3} \int d^3\bk$ we obtain

\begin{eqnarray}
\sum_{\bk j} \frac{\gkja \gkjb e^{\pm i\bk \cdot{\bf R}}} {\hbar(\wo+\wk)} & = & \frac{1}{\pi}\sum_{mn}d_m^A d_n^B D_{mn}^Rf(k_0R)
\end{eqnarray}
where the differential operator
\begin{eqnarray}
D_{mn}^R & \equiv & -\left( -\nabla^2\delta_{mn}+\nabla_m \nabla_n\right)^R \nonumber \\
& = & \frac{1}{R}\left[ \left( \delta_{mn}-\hat{R}_m\hat{R}_n
\right) \frac{\partial^2}{\partial R^2}+\left(
\delta_{mn}-3\hat{R}_m\hat{R}_n \right) \left(
\frac{1}{R^2}-\frac{1}{R} \frac{\partial}{\partial R} \right)
\right]
\end{eqnarray}
has been introduced for ease of notation and $f(x)$ denotes one of the auxiliary functions of the integral sine and cosine
functions \cite{Stegun}. Thus the concurrence takes the form

\begin{equation}
C=\frac{2}{\pi \hbar \wo} \; \left|\sum_{mn}d_m^A d_n^B D_{mn}^Rf(k_0R)\right| \label{concurr}\end{equation}

Using a terminology typical of the Casimir - Polder context, this
expression simplifies considerably when we examine its behavior in
the near zone, defined by $k_0R\ll 1$, and in the far zone,
defined by $k_0R \gg 1$:

\begin{eqnarray}
C_{\mbox{(near zone)}} &\simeq& \frac{|{\bf d}^{A} \cdot {\bf
d}^{B}-3 ({\bf d}^{A}\cdot \hat{{\bf R}})({\bf d}^{B}\cdot
\hat{{\bf R}}) |}{\hbar \wo R^3},
\;\;\;  k_0R \ll 1 \label{C near} \\
\nonumber\\
C_{\mbox{(far zone)}} &\simeq &\frac{8c \mid{\bf d}^{A} \cdot {\bf
d}^{B}-2 ({\bf d}^{A}\cdot \hat{{\bf R}})({\bf d}^{B}\cdot
\hat{{\bf R}}) \mid} {\pi \hbar \wo^2 R^4}, \;\;\; k_0R \gg 1
\label{C far}
\end{eqnarray}
The above expressions, which are the main result of this letter, are amenable of a straightforward physical interpretation. To
illustrate the relations between our expressions for the concurrence and the Casimir - Polder potential we first observe that this
latter can be written in the following form \cite{Lucia}:
\begin{equation}
\label{WCP} W_{C-P} = \Re \left( \sum_{\bk j, \ell m} \left[ \langle 0_k \mid E_{\bk j}({\bf R}_B )_m  E_{\bk j}({\bf R}_A)_\ell
\mid 0_k \rangle \alpha_A(k) \alpha_B(k) V_{\ell m}(k, \bR) \right] \right)
\end{equation}
where
\begin{equation}
\langle 0_k \mid E_{\bk j}({\bf R}_B )_m E_{\bk
j}({\bf R}_A)_\ell \mid 0_k \rangle =  \frac {2\pi \hbar c}V
\left( \ekj \right)_m \left( \ekj \right)_\ell k e^{i\bk \cdot
\bR} \label{eq:14}
\end{equation}
is the equal-time spatial correlation function of the electrical field modes in the vacuum state and evaluated at the position of
the two atoms ,

\begin{equation}
\alpha(k)=\frac{2\wo d^2}{3\hbar(\omega_{0}^2-\wk^2)}
\end{equation}
is the dynamic electric polarizability of the atoms and the quantity
\begin{eqnarray}
 V_{\ell m}(k, \bR ) &=& -D_{\ell m}^R \frac {\cos kR}R
\nonumber \\
&=& k^3 \left[ \left( \delta_{\ell m} -\hat{R}_\ell \hat{R}_m
\right) \frac {\cos kR}{kR} - \left( \delta_{\ell m}
-3\hat{R}_\ell \hat{R}_m \right) \left( \frac {\sin kR}{k^2R^2} +
\frac {\cos kR}{k^3R^3} \right) \right] \label{eq:15}
\end{eqnarray}
is the classical interaction  potential between two dipoles oscillating at frequency $ck$ \cite{MLP66}.  The form
(\ref{WCP}) of the Casimir--Polder energy emphasizes the role of the spatial correlations of the electromagnetic vacuum
state.

The link between the near - zone concurrence and the near zone Casimir - Polder $R^{-6}$ potential can be easily established by
noting that the latter coincides with the  well-known van der Waals potential between two neutral atoms. This was first derived by
London \cite{London} and can be obtained treating by second order perturbation theory the dipole - dipole interaction hamiltonian
\cite{Cohen}
\begin{equation}
\label{ }
\frac{ {\bf  d}^{A} \cdot {\bf
d}^{B}-3 ({\bf d}^{A}\cdot \hat{{\bf R}})({\bf d}^{B}\cdot
\hat{{\bf R}})  }{R^3},
\end{equation}
In the near - zone the Casimir -Polder potential is therefore essentially of electrostatic nature. Note that indeed that the
expression (\ref{C near}) for the concurrence in the near zone,  is the ratio between the interaction energy between two permanent
dipoles and the energy separation $\hbar\omega_0$.

The fact that in the near - zone the interaction between the two atoms is essentially of electrostatic nature and can be described
by an effective Hamiltonian in which the field degrees of freedom are eliminated suggests that in such region the state of the two
atoms is described by a pure density operator. Indeed, in the near zone, we have $\rho_{AB}^2=\rho_{AB}$ up to second-order
approximation if only the leading terms in the expansion in series of powers of $k_0R$ are kept. This implies that the field
induces pure bipartite entanglement between the probe atoms i.e. there is no entanglement between atoms and field but only
entanglement between the atoms which is mediated by the field.

 The "far - zone" $R^{-7}$ behavior of the Casimir-Polder
potential, eq.(\ref{WCP}), stems from retardation effects, and it is a typical manifestation of the quantum nature of the
electromagnetic field. In an analogous way, a change of the power law ($R^{-4}$ instead of $R^{-3}$) is found in the concurrence
when we move to the far zone (see eq.(\ref{C far})). Here entanglement can be interpreted as a consequence of vacuum correlations
of the electric field ${\bf E}$ \cite{biswas}, since the concurrence can be cast in the form

\begin{equation}
\label{Cfar} C_{\mbox{(far zone)}} =\sum_{m n}\frac{2 d^A_m  d^B_n \mid \langle E_{m}({\bf R}_A)E_n({\bf R}_B) \rangle
\mid}{\hbar^2 \wo^2}, \;\;\;\; k_0R \gg 1
\end{equation}
The analogy with the Casimir-Polder potential becomes more transparent if we note that eq. (\ref{WCP} ), in the far zone, can be
cast in the following form:"

\begin{equation}
\label{Wfar} W_{C-P\mbox{(far zone)}} = \sum_{m n i\ell } \frac{2 d^A_i d^A_m d^B_\ell d^B_n }{9\hbar^2 \wo^2} \sum_{\bk j}
\langle E_{\bk j}({\bf R}_B )_m E_{\bk j}({\bf R}_A)_n \rangle V_{i \ell}(k, \bR)
\end{equation}
Both (\ref{Cfar}) and (\ref{Wfar}) show clearly the role of the spatial correlations of the quantum field vacuum fluctuations in
correlating the atomic dipoles.  The only difference is that while the concurrence (\ref{Cfar}) is due to spatial correlations
between the fields at positions ${\bf R}_A$ and ${\bf R}_B$ while the Casimir - Polder potential (\ref{Wfar}) is due to the
correlations between the field {\it modes } at positions ${\bf R}_A$ and ${\bf R}_B$. Note incidentally that in the far  zone the
system (atoms plus field) manifests an essentially tripartite entanglement, since in this region we have $\rho^2_{AB}\neq
\rho_{AB}$.

In summary we have evaluated the entanglement between two spatially separated two-level atoms interacting with the vacuum state of
the radiation field. In particular we have shown that the concurrence can be cast in a form with strong analogies with the Casimir
- Polder potential.  The same experimental setup used to measure such potential can be used to infer the amount of entanglement
between  two spatially separated neutral atoms.  To estimate the expected order of magnitude of the concurrence one can assume
that the two atoms are two hydrogen atoms. This gives in the near zone $ C_{\mbox{n.z.}} \approx (R/a_0)^{-3}$ where $a_0$ is the
Bohr radius, while in the far zone $C_{\mbox{f.z.}} \approx \alpha (R/a_0)^{-4}$ where $\alpha = e^2/(\hbar c) $ is the
dimensionless fine structure constant.

The above discussion provides a physically transparent characterization of the entanglement created by the non local zero-point field fluctuations and suggests a strategy for its experimental detection.

\acknowledgments
 M.C and G.M.P. acknowledge financial support under PRIN 2006 "Quantum noise in mesoscopic systems"

\end{document}